# The Statistical Analysis of fMRI Data

**Martin A. Lindquist**

*Abstract.*   In recent years there has been explosive growth in the number of neuroimaging studies performed using functional Magnetic Resonance Imaging (fMRI). The field that has grown around the acquisition and analysis of fMRI data is intrinsically interdisciplinary in nature and involves contributions from researchers in neuroscience, psychology, physics and statistics, among others. A standard fMRI study gives rise to massive amounts of noisy data with a complicated spatio-temporal correlation structure. Statistics plays a crucial role in understanding the nature of the data and obtaining relevant results that can be used and interpreted by neuroscientists. In this paper we discuss the analysis of fMRI data, from the initial acquisition of the raw data to its use in locating brain activity, making inference about brain connectivity and predictions about psychological or disease states. Along the way, we illustrate interesting and important issues where statistics already plays a crucial role. We also seek to illustrate areas where statistics has perhaps been underutilized and will have an increased role in the future.

## 1. INTRODUCTION

Functional neuroimaging has experienced an explosive growth in recent years. Currently there exist a number of different imaging modalities that allow researchers to study the physiological changes that accompany brain activation. Each of these techniques has advantages and disadvantages and each provides a unique perspective on brain function. In general, these techniques are not concerned with the behavior of single neurons, but rather with activity arising from a large group of neurons. However, they differ in what they attempt to measure, as well as in the temporal and spatial resolution that they provide. Techniques such as electroencephalography (EEG) and magnetoencephalography (MEG) are based on studying electrical and magnetic activity in the brain. They provide temporal resolution on the order of milliseconds but uncertain spatial localization. In contrast, functional magnetic resonance imaging (fMRI) and positron emission tomography (PET) provide information on blood flow changes that accompany neuronal activity with relatively high spatial resolution, but with a temporal resolution limited by the much slower rate of brain hemodynamics. While each modality is interesting in its own right, this article focuses on statistical issues related to fMRI which in the past few years has taken a dominant position in the field of neuroimaging.

Functional MRI is a noninvasive technique for studying brain activity. During the course of an fMRI experiment, a series of brain images are acquired while the subject performs a set of tasks. Changes in the measured signal between individual images are used to make inferences regarding task-related activations in the brain. fMRI has provided researchers with unprecedented access to the brain in action and, in the past decade, has provided countless new insights into the inner workings of the human brain.

*Martin A. Lindquist is Associate Professor, Department of Statistics, Columbia University, New York, New York 10027, USA e-mail: martin@stat.columbia.edu.*

There are several common objectives in the analysis of fMRI data. These include localizing regions of the brain activated by a task, determining distributed networks that correspond to brain function and making predictions about psychological or disease states. Each of these objectives can be approached through the application of suitable statistical methods, and statisticians play an important role in the interdisciplinary teams that have been assembled to tackle these problems. This role can range from determining the appropriate statistical method to apply to a data set, to the development of unique statistical methods geared specifically toward the analysis of fMRI data. With the advent of more sophisticated experimental designs and imaging techniques, the role of statisticians promises to only increase in the future.

The statistical analysis of fMRI data is challenging. The data comprise a sequence of magnetic resonance images (MRI), each consisting of a number of uniformly spaced volume elements, or voxels, that partition the brain into equally sized boxes. The image intensity from each voxel represents the spatial distribution of the nuclear spin density in that area. Changes in brain hemodynamics, in reaction to neuronal activity, impact the local intensity of the MR signal, and therefore changes in voxel intensity across time can be used to infer when and where activity is taking place.

During the course of an fMRI experiment, images of this type are acquired between 100–2000 times, with each image consisting of roughly 100,000 voxels. Further, the experiment may be repeated several times for the same subject, as well as for multiple subjects (typically between 10–40) to facilitate population inference. Though a good number of these voxels consist solely of background noise, and can be excluded from further analysis, the total amount of data that needs to be analyzed is staggering. In addition, the data exhibit a complicated temporal and spatial noise structure with a relatively weak signal. A full spatiotemporal model of the data is generally not considered feasible and a number of short cuts are taken throughout the course of the analysis. Statisticians play an important role in determining which short cuts are appropriate in the various stages of the analysis, and determining their

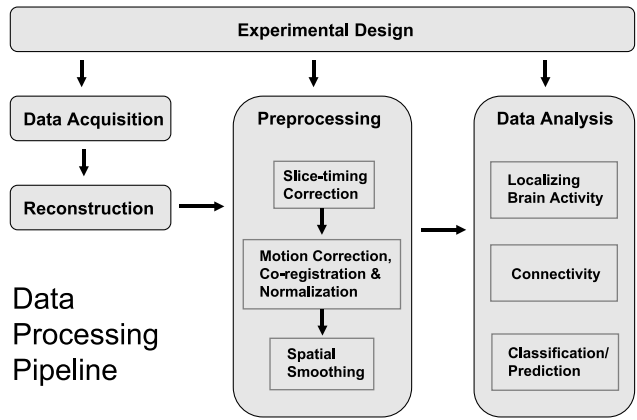

FIG. 1. *The fMRI data processing pipeline illustrates the different steps involved in a standard fMRI experiment. The pipeline shows the path from the initial experimental design to the acquisition and reconstruction of the data, to its preprocessing and analysis. Each step in the pipeline contains interesting mathematical and statistical problems.*

effects on the validity and power of the statistical analysis.

fMRI has experienced a rapid growth in the past several years and has found applications in a wide variety of fields, such as neuroscience, psychology, economics and political science. This has given rise to a bounty of interesting and important statistical problems that cover a variety of topics, including the acquisition of raw data in the MR scanner, image reconstruction, experimental design, data preprocessing and data analysis. Figure 1 illustrates the steps involved in the data processing pipeline that accompanies a standard fMRI experiment. To date, the primary domain of statisticians in the field has been the data analysis stage of the pipeline, though many interesting statistical problems can also be found in the other steps. In this paper we will discuss each step of the pipeline and illustrate the important role that statistics plays, or can play. We conclude the paper by discussing a number of additional statistical challenges that promise to provide important areas of research for statisticians in the future.

## 2. ACQUIRING fMRI DATA

The data collected during an fMRI experiment consists of a sequence of individual magnetic resonance images, acquired in a manner that allows one to study oxygenation patterns in the brain. Therefore, to understand the nature of fMRI data and how these images are used to infer neuronal activity, one must first study the acquisition of individual MR images. The overview presented here is by necessity brief and we refer interested readers to any number of introductory text books (e.g., Haacke et al., 1999) dealing specifically with MR physics. In addition, it will also be critical in subsequent data analysis to have a clear understanding of the statistical properties of the resulting images, and their distributional properties will be discussed. Finally, we conclude with a brief discussion linking MRI to fMRI.

## 2.1 Data Acquisition

To construct an image, the subject is placed into the field of a large electromagnet. The magnet has a very strong magnetic field, typically between 1.5–7.0 Tesla,[1] which aligns the magnetization of hydrogen ($^1H$) atoms in the brain. Within a slice of the brain, a radio frequency pulse is used to tip over the aligned nuclei. Upon removal of this pulse, the nuclei strive to return to their original aligned positions and thereby induce a current in a receiver coil. This current provides the basic MR signal. A system of gradient coils is used to sequentially control the spatial inhomogeneity of the magnetic field, so that each measurement of the signal can be approximately expressed as the Fourier transformation of the spin density at a single point in the frequency domain, or $k$-space as it is commonly called in the field. Mathematically, the measurement of the MR signal at the $j$th time point of a readout period can be written

$$S(t_j) \approx \int_x \int_y M(x,y) \cdot e^{(-2\pi i(k_x(t_j)x + k_y(t_j)y))} \, dx \, dy, \quad (1)$$

where $M(x,y)$ is the spin density at the point $(x,y)$, and $(k_x(t_j), k_y(t_j))$ is the point in the frequency domain ($k$-space) at which the Fourier transformation is measured at time $t_j$. Here $t_j = j\Delta_t$ is the time of the $j$th measurement, where $\Delta_t$ depends on the sampling bandwidth of the scanner; typically it takes values in the range of 250–1000 $\mu s$.

To reconstruct a single MR image, one needs to sample a large number of individual $k$-space measurements, the exact number depending on the desired image resolution. For example, to fully reconstruct a $64 \times 64$ image, a total of 4096 separate measurements are required, each sampled at a unique coordinate of $k$-space. There is a time cost involved in sampling each point, and therefore the time it takes to acquire an image is directly related to its spatial resolution. There are a variety of approaches toward sampling the data. Traditionally, it has been performed on a Cartesian grid which is uniformly spaced and symmetric about the origin of $k$-space. This method of sampling, as illustrated in Figure 2, allows for the quick and easy reconstruction of the image using the Fast Fourier Transform (FFT). Recently, it has become increasingly popular to sample $k$-space using nonuniform trajectories; a particularly popular trajectory has been the Archimedean spiral (Glover, 1999b). While such trajectories provide a number of benefits relating to speed and signal-to-noise ratio, the FFT algorithm cannot be directly applied to the nonuniformly sampled raw data. As a solution to this problem, the raw data are typically interpolated onto a Cartesian grid in $k$-space and thereafter the FFT is applied to reconstruct the image (Jackson et al., 1991).

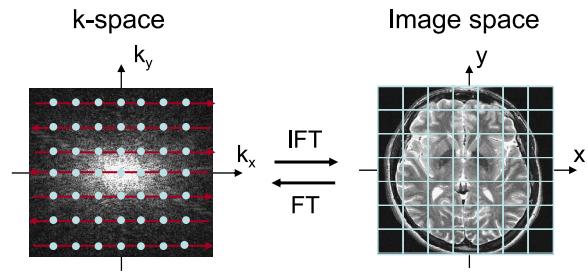

While the description so far has focused on sampling a single two-dimensional (2D) slice of the brain, most studies require the acquisition of a full 3D brain volume. The standard approach toward 3D imaging is to acquire a stack of adjacent slices (e.g., 20–30) in quick succession. Since the nuclei must be re-excited before sampling a new slice, this places constraints on the time needed to acquire a brain volume. Using this methodology, it takes approximately 2 seconds to obtain a full brain volume of dimension $64 \times 64 \times 30$. As an alternative, it is possible to design a sampling trajectory that directly samples points in 3D $k$-space (Mansfield, Howseman and Ordidge, 1989; Mansfield, Coxon and Hykin, 1995; Lindquist et al., 2008b). Though this approach would potentially allow the same number of $k$-space points to be sampled at a faster rate, the stacked slice approach remains dominant. However, with increases in computational power and hardware improvements, 3D sampling should attract increased attention.

The process of designing new $k$-space sampling trajectories is an interesting mathematical problem, which can easily be generalized to three dimensions by letting $\mathbf{k}(t) = (k_x(t), k_y(t)), k_z(t))$. The goal is to find a trajectory $\mathbf{k}(t)$ that moves through $k$-space and satisfies the necessary constraints. The trajectory is defined as a continuous curve and along this curve measurements are made at uniform time intervals determined by the sampling bandwidth of the scanner. The trajectory starts at the point $(0, 0, 0)$ and its subsequent movement is limited by constraints placed on both its speed and acceleration. In addition, there is a finite amount of time the signal can be measured before the nuclei need to be re-exited and the trajectory is returned to the origin. Finally, the trajectory needs to be space-filling, which implies that each point in the lattice contained within some cubic or spherical region around the center of $k$-space needs to be visited long enough to make a measurement. The size of this region determines the spatial resolution of the subsequent image reconstruction. For a more complete formulation of the problem, see Lindquist et al. (2008a). The problem bears some resemblance to the traveling salesman problem and can be approached in an analogous manner. One application where trajectory design is important is rapid imaging (Lindquist et al., 2006, 2008a) and we return to this issue in a later section.

### 2.2 Statistical Properties of MR Images

As the signal in (1) is measured over two channels, the raw $k$-space data are complex valued. It is assumed that both the real and imaginary component is measured with independent normally distributed error. Since the Fourier transformation is a linear operation, the reconstructed voxel data will also be complex-valued with both parts following a normal distribution. In the final stage of the reconstruction process, these complex valued measurements are separated into magnitude and phase components. In the vast majority of studies only the magnitude portion of the signal is used in the data analysis, while the phase portion is discarded. Traditionally, the phase has not been considered to contain relevant signal information, though models that use both components (Rowe and Logan, 2004) have been proposed. It should be noted that the magnitude values no longer follow a normal distribution, but rather a Rice distribution (Gudbjartsson and Patz, 1995). The shape of this distribution depends on the signal-to-noise (SNR) ratio within the voxel. For the special case when no signal is present (e.g., for voxels outside of the brain), it behaves like a Rayleigh distribution. When the SNR is high (e.g., for voxels within the brain) the distribution is approximately Gaussian. Understanding the distributional properties of MR images is important, and this area provides some interesting research opportunities for statisticians in terms of developing methods for estimating the variance of the background noise and methods for identifying and removing outliers that arise due to acquisition artifacts.

### 2.3 From MRI to fMRI

The data acquisition and reconstruction techniques outlined in this section provide the means for obtaining a static image of the brain. However, changes in brain hemodynamics in response to neuronal activity impact the local intensity of the MR signal. Therefore, a sequence of properly acquired brain images allows one to study changes in brain function over time.

An fMRI study consists of a series of brain volumes collected in quick succession. The temporal resolution of the acquired data will depend on the time between acquisitions of each individual volume; once the $k$-space has been sampled, the procedure is ready to be repeated and a new volume can be acquired. This is one reason why efficient sampling of $k$-space is important. Typically, brain volumes of dimensions $64 \times 64 \times 30$ (i.e. 122,880 voxels) are collected at $T$ separate time points throughout the course of an experiment, where $T$ varies between 100–2000. Hence, the resulting data consists of roughly 100,000 time series of length $T$. On top of this, the experiment is often repeated for $M$ subjects, where $M$ usually varies between 10 and 40. It quickly becomes clear that fMRI data analysis is a time series analysis problem of massive proportions.

## 3. UNDERSTANDING fMRI DATA

The ability to connect the measures of brain physiology obtained in an fMRI experiment with the underlying neuronal activity that caused them will greatly impact the choice of inference procedure and the subsequent conclusions that can be made. Therefore, it is important to gain some rudimentary understanding of basic brain physiology. The overview presented here is brief and interested readers are referred to text books dealing specifically with the subject (e.g., Huettel, Song and Mccarthy, 2004). In addition, since neuronal activity unfolds both in space and time, the spatial and temporal resolution of fMRI studies will limit any conclusions that can be made from analyzing the data and understanding these limitations is paramount. Finally, as relatively small changes in brain activity are buried within noisy measurements, it will be important to understand the behavior of both the signal and noise present in fMRI data and begin discussing how these components can be appropriately modeled.

### 3.1 BOLD fMRI

Functional magnetic resonance imaging is most commonly performed using blood oxygenation level-dependent (BOLD) contrast (Ogawa et al., 1992) to study local changes in deoxyhemoglobin concentration in the brain. BOLD imaging takes advantage of inherent differences between oxygenated and deoxygenated hemoglobin. Each of these states has different magnetic properties, diamagnetic and paramagnetic respectively, and produces different local magnetic fields. Due to its paramagnetic properties, deoxy-hemoglobin has the effect of suppressing the MR signal, while oxy-hemoglobin does not. The cerebral blood flow refreshes areas of the brain that are active during the execution of a mental task with oxygenated blood, thereby changing the local magnetic susceptibility and the measured MR signal in active brain regions. A series of properly acquired MR images can therefore be used to study changes in blood oxygenation which, in turn, can be used to infer brain activity.

The underlying evoked hemodynamic response to a neural event is typically referred to as the hemodynamic response function (HRF). Figure 3A shows the standard shape used to model the HRF, sometimes called the canonical HRF. The increased metabolic demands due to neuronal activity lead to an increase in the inflow of oxygenated blood to active regions of the brain. Since more oxygen is supplied than actually consumed, this leads to a decrease in the concentration of deoxy-hemoglobin which, in turn, leads to an increase in signal. This positive rise in signal has an onset approximately 2 seconds after the onset of neural activity and peaks 5–8 seconds after that neural activity has peaked (Aguirre, Zarahn and D'Esposito, 1998). After reaching its peak level, the BOLD signal decreases to a below baseline level which is sustained for roughly 10 seconds. This effect, known as the post-stimulus undershoot, is due to the fact that blood flow decreases more rapidly than blood volume, thereby allowing for a greater concentration of deoxy-hemoglobin in previously active brain regions.

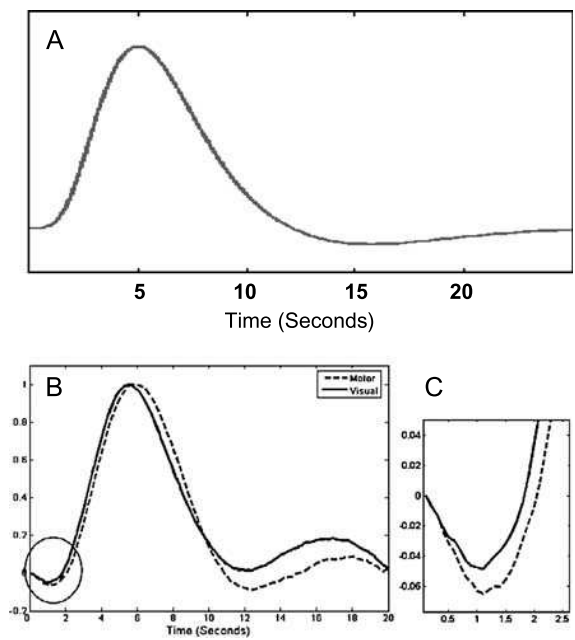

Fig. 3. (A) *The standard canonical model for the HRF used in fMRI data analysis illustrates the main features of the response.* (B) *Examples of empirical HRFs measured over the visual and motor cortices in response to a visual-motor task.* (C) *The initial 2 seconds of the empirical HRFs give strong indication of an initial decrease in signal immediately following activation.*

Several studies have shown evidence of a decrease in oxygenation levels in the time immediately following neural activity, giving rise to a decrease in the BOLD signal in the first 1–2 seconds following activation. This decrease is called the initial negative BOLD response or the negative dip (Menon et al., 1995; Malonek and Grinvald, 1996). Figures 3B–C illustrate this effect in data collected during an experiment that stimulated both the visual and motor cortices. The ratio of the amplitude of the dip compared to the positive BOLD signal depends on the strength of the magnet and has been reported to be roughly 20% at 3 Tesla (Yacoub, Le and Hu, 1998). There is also evidence that the dip is more localized to areas of neural activity (Yacoub, Le and Hu, 1998; Duong et al., 2000; Kim, Duong and Kim, 2000; Thompson, Peterson and Freeman, 2004) than the subsequent rise which appears less spatially specific. Due in part to these reasons, the negative response has so far not been reliably observed and its existence remains controversial (Logothetis, 2000).

### 3.2 Spatial and Temporal Limitations

There are a number of limitations that restrict what fMRI can measure and what can be inferred from an fMRI study. Many of these limitations are directly linked to the spatial and temporal resolution of the study. When designing an experiment it is therefore important to balance the need for adequate spatial resolution with that of adequate temporal resolution. The temporal resolution determines our ability to separate brain events in time, while the spatial resolution determines our ability to distinguish changes in an image across spatial locations. The manner in which fMRI data is collected makes it impossible to simultaneously increase both, as increases in temporal resolution limit the number of $k$-space measurements that can be made in the allocated sampling window and thereby directly influence the spatial resolution of the image. Therefore, there are inherent trade-offs required when determining the appropriate spatial and temporal resolutions to use in an fMRI experiment.

One of the benefits of MRI as an imaging technique is its ability to provide detailed anatomical scans of gray and white matter with a spatial resolution well below 1 mm$^3$. However, the time needed to acquire such scans is prohibitively high and currently not feasible for use in functional studies. Instead, the spatial resolution is typically on the order of $3 \times 3 \times 5$ mm$^3$, corresponding to image dimensions on the order of $64 \times 64 \times 30$, which can readily be sampled in approximately 2 seconds. Still, fMRI provides relatively high spatial resolution compared with many other functional imaging techniques. However, it is important to note that the potential high spatial resolution is often limited by a number of factors. First, it is common to spatially smooth fMRI data prior to analysis which decreases the effective resolution of the data. Second, performing population inference requires the analysis of groups of subjects with varying brain sizes and shapes. In order to compare data across subjects, a normalization procedure is used to warp the brains onto a standard template brain. This procedure introduces spatial imprecision and blurring in the group data. An obvious impact of all this blurring is that activation in small structures may be mislocalized or even missed all together.

Inferences in space can potentially be improved by advances in data acquisition and preprocessing. The introduction of enhanced spatial inter-subject normalization techniques and improved smoothing techniques would help researchers avoid the most dramatic effects of blurring the data. Statistical issues that arise due to smoothing and normalization will be revisited in a later section dealing specifically with preprocessing. A recent innovation in signal acquisition has been the use of multiple coils with different spatial sensitivities to simultaneously measure $k$-space (Sodickson and Manning, 1997; Pruessmann et al., 1999). This approach, known as parallel imaging, allows for an increase in the amount of data that can be collected in a given time window. Hence, it can be used to either increase the spatial resolution of an image or decrease the amount of time required to sample an image with a certain specified spatial resolution. Parallel imaging techniques have already had a great influence on the way data is collected and its role will only increase. The appropriate manner to deal with parallel imaging data is a key direction for future research. Designing new ways of acquiring and reconstructing multi-coil data is an important area of research where statistics can play a vital role.

The temporal resolution of an fMRI study depends on the time between acquisition of each individual image, or the repetition time (TR). In most fMRI studies the TR ranges from 0.5–4.0 seconds. These values indicate a fundamental disconnect between the underlying neuronal activity, which takes place on the order of tens of milliseconds, and the temporal resolution of the study. However, the statistical analysis of fMRI data is primarily focused on using the positive BOLD response to study the underlying neural activity. Hence, the limiting factor in determining the appropriate temporal resolution is generally not considered the speed of data acquisition, but rather the speed of the underlying evoked hemodynamic response to a neural event. Since inference is based on oxygenation patterns taking place 5–8 seconds after activation, TR values in the range of 2 seconds are generally deemed adequate.

Because of the relatively low temporal resolution and the sluggish nature of the hemodynamic response, inference regarding when and where activation is taking place is based on oxygenation patterns outside of the immediate vicinity of the underlying event we want to base our conclusions on (i.e., the neural activity). Since the time-to-peak positive BOLD response occurs in a larger time scale than the speed of brain operations, there is a risk of unknown confounding factors influencing the ordering of time-to-peak relative to the ordering of brain activation in different regions of interest. For these reasons it is difficult to determine the *absolute* timing of brain activity using fMRI. However, studies have shown (Menon, Luknowsky and Gati, 1998; Miezin et al., 2000) that the *relative* timing within a voxel in response to different stimuli can be accurately captured. There are also indications that focusing inference on features related to the initial dip can help alleviate concerns (Lindquist et al., 2008a) regarding possible confounders. However, these types of studies require significant increases in temporal resolution and the ability to rapidly acquire data becomes increasingly important. Finally, another way of improving inferences in time is through appropriate experimental design. In principal, it is possible to estimate the HRF at a finer temporal resolution than the TR as long as the onsets of repeated stimuli are jittered in time (Dale, 1999). For example, if the onset is shifted by TR/2 in half of the stimuli, one can potentially estimate the HRF at a temporal resolution of TR/2, rather than TR if jittering is not used.

### 3.3 Understanding Signal and Noise

Prior to modeling fMRI data, it is useful to gain a better understanding of the components present in an fMRI time series. In general, it consists of the BOLD signal (which is the component of interest), a number of nuisance parameters and noise. Each component is discussed in detail below.

*BOLD signal.* The evoked BOLD response in fMRI is a complex, nonlinear function of the results of neuronal and vascular changes (Wager et al., 2005), complicating the ability to appropriately model its behavior. The shape of the response depends both on the applied stimulus and the hemodynamic response to neuronal events. A number of methods for modeling the BOLD response and the underlying HRF exist in the literature. A major difference between methods lies in how the relationship between the stimulus and BOLD response is modeled. We differentiate between nonlinear physiological-based models, such as the Balloon model (Buxton, Wong and Frank, 1998; Friston et al., 2000; Riera et al., 2004), which describes the dynamics of cerebral blood volume and deoxygenation and their effects on the resulting BOLD signal, and models that assume a linear time invariant (LTI) system, in which assumed neuronal activity (based on task manipulations) constitutes the input, or impulse, and the HRF is the impulse response function.

The Balloon model consists of a set of ordinary differential equations that model changes in blood volume, blood inflow, deoxyhemoglobin and flow-inducing signal and how these changes impact the observed BOLD response. While models of this type tend to be more biophysically plausible than their linear counterparts, they have a number of drawbacks. First, they require the estimation of a large number of model parameters. Second, they do not always provide reliable estimates with noisy data, and third, they do not provide a direct framework for performing inference. In general, they are not yet considered feasible alternatives for performing whole-brain multi-subject analysis of fMRI data in cognitive neuroscience studies, although promising developments are being pursued on this front and this is an exciting area for future research.

While the flexibility of nonlinear models is attractive, linear models provide robust and interpretable characterizations in noisy systems. It is therefore common to assume a linear relationship between neuronal activity and BOLD response, where linearity implies that the magnitude and shape of the evoked HRF do not depend on any of the preceding stimuli. Studies have shown that under certain conditions the BOLD response can be considered linear with respect to the stimulus (Boynton et al., 1996), particularly if events are spaced at least 5 seconds apart (Miezin et al., 2000). However, other studies have found that nonlinear effects in rapid sequences (e.g., stimuli spaced less than 2 seconds apart) can be quite large (Birn, Saad and Bandettini, 2001; Wager et al., 2005). The ability to assume linearity is important, as it allows the relationship between stimuli and the BOLD response to be modeled using a linear time invariant system, where the stimulus acts as the input and the HRF acts as the impulse response function. Figure 4 shows an illustration of the estimated BOLD signal corresponding to two different types of stimulus patterns. In a linear system framework the signal at time $t$, $x(t)$, is modeled as the convolution of a stimulus function $v(t)$ and the hemodynamic response $h(t)$, that is,

$$(2) \qquad x(t) = (v * h)(t).$$

Here $h(t)$ is either assumed to take a canonical form, or alternatively modeled using a set of linear basis functions.

Another important modeling aspect is that the timing and shape of the HRF are known to vary across the brain, within an individual and across individuals (Aguirre, Zarahn and D'Esposito, 1998; Schacter et al., 1997). Part of the variability is due to the underlying configuration of the vascular bed, which may cause differences in the HRF across brain regions in the same task for purely physiological reasons (Vazquez et al., 2006). Another source of variability is differences in the pattern of evoked neural activity in regions performing different functions related to the same task. It is important that these regional variations are accounted for when modeling the BOLD signal and we return to this issue in a later section.

In general, one of the major shortfalls when analyzing fMRI data is that users typically assume a canonical HRF (Grinband et al., 2008), which leaves open the possibility for mismodeling the signal in large portions of the brain (Loh, Lindquist and Wager, 2008). There has therefore been a movement toward both using more sophisticated models and enhanced model diagnostics. Both of these areas fall squarely in the purview of statisticians, and promise to have increased importance in the future.

*Noise and nuisance signal.* The measured fMRI signal is corrupted by random noise and various nuisance components that arise due both to hardware reasons and the subjects themselves. For instance, fluctuations in the MR signal intensity caused by thermal motion of electrons within the subject and the scanner gives rise to noise that tends to be highly random and independent of the experimental task. The amount of thermal noise increases linearly as a function of the field strength of the scanner, with higher field strengths giving rise to more noise. However, it does not exhibit spatial structure and its effects can be minimized by averaging the signal over multiple data points. Another source of variability in the signal is due to scanner drift, caused by scanner instabilities, which result in slow changes in voxel intensity over time (low-frequency noise). The amount of drift varies across space, and it is important to include this source of variation in your models. Finally, physiological noise due to patient motion, respiration and heartbeat cause fluctuations in signal across both space and time. Physiological noise can often be modeled and the worst of its effects removed. In the next section we discuss how to correct for subject motion as part of the preprocessing step of the analysis. However, heart-rate and respiration gives rise to periodic fluctuations that are difficult to model. According to the Nyquist criteria, it is necessary to have a sampling rate at least twice as high as the frequency of the periodic function one seeks to model. If the TR is too low, which is true in most fMRI studies, there will be problems with aliasing; see Figure 5A for an illustration. In this situation the periodic fluctuations will be distributed throughout the time course giving rise to temporal autocorrelation. Noise in fMRI is typically modeled using either an AR($p$) or an ARMA(1, 1) process (Purdon et al., 2001), where the autocorrelation is thought to be due to an unmodeled nuisance signal. If these terms are properly removed, there is evidence that the resulting error term corresponds to white noise (Lund et al., 2006). Note that for high temporal resolution studies, heart-rate and respiration can be estimated and included in the model, or alternatively removed through application of a band-pass filter.

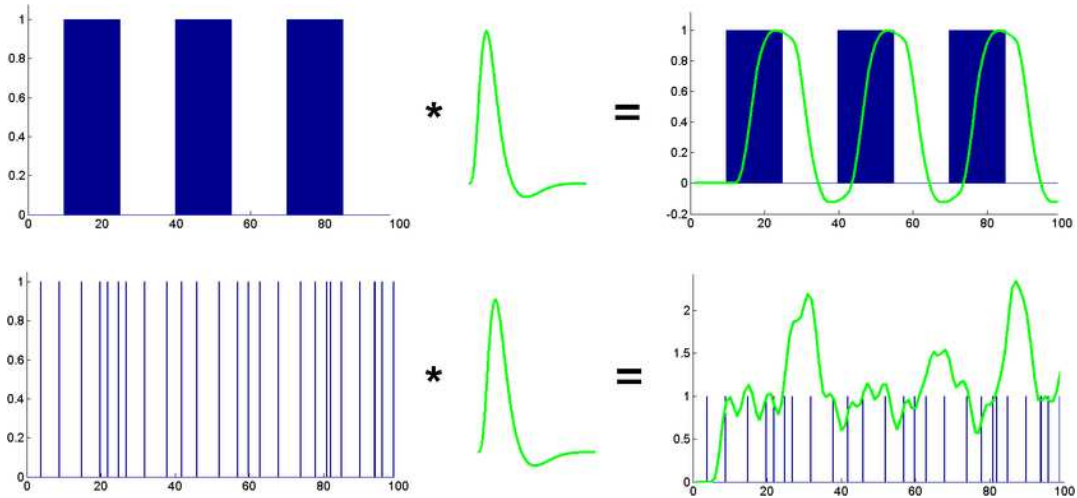

FIG. 4. *The BOLD response is typically modeled as the convolution of the stimulus function with the HRF. Varying stimulus patterns will give rise to responses with radically different features.*

The spatiotemporal behavior of the noise process is complex. Figure 5B shows a time course from a single voxel sampled at high temporal resolution (60 ms), as well as its power spectrum. The power spectrum indicates periodic oscillations in the signal due to physiological effects and a low-frequency component corresponding to signal drift. At this resolution it is relatively straightforward to remove the effects of these nuisance functions by applying an appropriate filter. In contrast, Figure 5C shows a time course sampled at a more standard resolution (1 s). At this resolution, the sampling rate is too low to effectively model physiological noise and it gives rise to temporal autocorrelation clearly visible in the accompanying autocorrelation plot. Finally, Figure 5D shows spatial maps of the model parameters from an AR(2) model estimated for each voxel's noise data. It is clear that the behavior of the noise is not consistent throughout the image, indicating spatial dependence. In fact, it is clearly possible to make out rough anatomical detail in the maps, indicating higher amounts of variability in certain brain regions.

## 4. EXPERIMENTAL DESIGN

The experimental design of an fMRI study is complicated, as it not only involves the standard issues relevant to psychological experiments, but also issues related to data acquisition and stimulus presentation. Not all designs with the same number of trials of a given set of conditions are equal, and the spacing and ordering of events is critical. What constitutes an optimal experimental design depends on the psychological nature of the task, the ability of the fMRI signal to track changes introduced by the task over time and the specific comparisons that one is interested in making. In addition, as the efficiency of the subsequent statistical analysis is directly related to the experimental design, it is important that it be carefully considered during the design process.

A good experimental design attempts to maximize both statistical power and psychological validity. The statistical performance can be characterized by its estimation efficiency (i.e., the ability to estimate the HRF) and its detection power (i.e., the ability to detect significant activation). The psychological validity is often measured by the randomness of the stimulus presentation, as this helps control for issues related to anticipation, habituation and boredom. When designing an experiment there is inherent trade-offs between estimation efficiency, detection power and randomness. The optimal balance between the three ultimately depends on the goals of the experiment and the combination of conditions one is interested in studying. For example, a design used to localize areas of brain activation stresses high detection power at the expensive of estimation efficiency and randomness.

While the area of experimental design is a natural domain for statisticians to conduct research, it has so far been largely unexplored by members of the field. Currently there are two major classes of fMRI experimental designs: block designs and event-related designs. In the following sections we describe each type and discuss the applications for which they are best suited. In addition, we also discuss ways of optimizing the experimental design.

### Block Designs

In a block design the different experimental conditions are separated into extended time intervals, or blocks. For example, one might repeat the process of interest (e.g., finger tapping) during an experimental block (A) and have the subject rest during a control block (B); see Figure 6. The $A$–$B$ comparison can than be used to compare differences in signal between the conditions. In general, increasing the length of each block will lead to a larger evoked response during the task. This increases the separation in signal between blocks, which, in turn, leads to higher detection power. However, in contrast, it is also important to include multiple transitions between conditions, as otherwise differences in signal due to low-frequency drift may be confused for differences in task conditions. In addition, it is important that the same mental processes are evoked throughout each block. If block lengths are too long, this assumption may be violated due to the effects of fatigue and/or boredom.

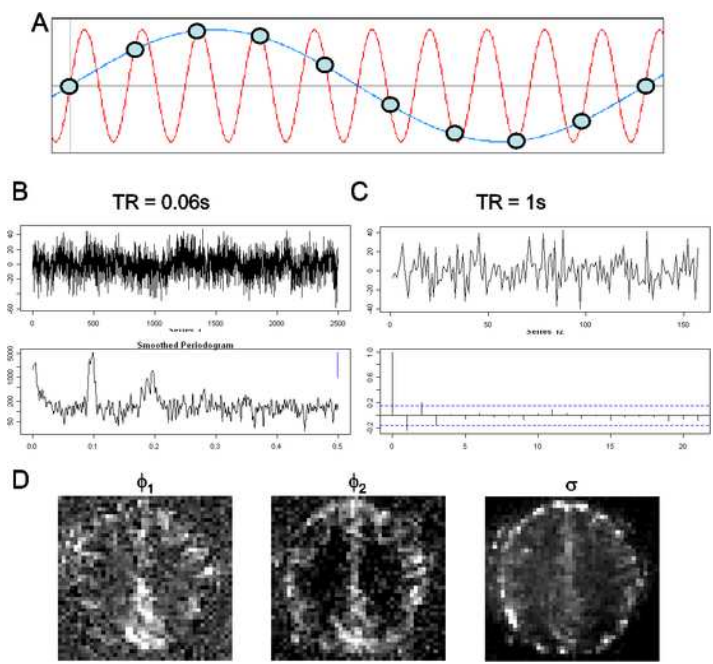

FIG. 5.  (A) *The Nyquist criteria states that it is necessary to sample at a frequency at least twice as high as the frequency of the periodic function one seeks to model to avoid aliasing. As an illustration assume that the signal is measured at the time points indicated by circles. In this situation it is impossible to determine which of the two periodic signals shown in the plot give rise to the observed measurements.* (B) *An fMRI time course measured at a single voxel sampled with 60 ms resolution. Its power spectrum indicates components present in the signal whose periodicity corresponds to low-frequency drift and physiological effects.* (C) *An fMRI time course measured with 1 s resolution. The autocorrelation function indicates autocorrelation present in the signal.* (D) *Spatial maps of the model parameters from an AR(2) model (i.e. $\phi_1$, $\phi_2$ and $\sigma$), estimated from each voxel's noise data, indicates clear spatial dependence.*

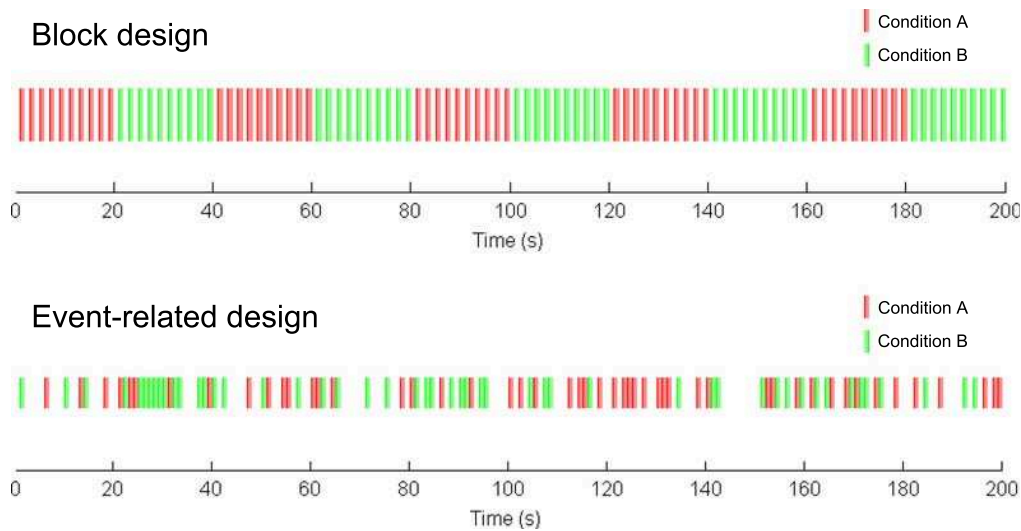

FIG. 6.  *The two most common classes of experimental design are block designs and event-related designs. In a block design (top) experimental conditions are separated into extended time intervals, or blocks, of the same type. In an event-related design (bottom) the stimulus consists of short discrete events whose timing and order can be randomized.*

The main advantages to using a block design are that they offer high statistical power to detect activation and are robust to uncertainties in the shape of the HRF. The latter advantage is due to the fact that the predicted response depends on the total activation caused by a series of stimuli, which makes it less sensitive to variations in the shape of responses to individual stimulus (see Figure 4). The flip side is that block designs provide imprecise information about the particular processes that activated a brain region and cannot be used to directly estimate important features of the HRF (e.g., onset or width).

### Event-Related Designs

In an event-related design the stimulus consists of short discrete events (e.g., brief light flashes) whose timing can be randomized; see Figure 6 for an illustration with two conditions. These types of designs are attractive because of their flexibility and that they allow for the estimation of key features of the HRF (e.g., onset and width) that can be used to make inference about the relative timing of activation across conditions and about sustained activity. Event-related designs allow one to discriminate the effects of different conditions as long as one either intermixes events of different types or varies the inter-stimulus interval between trials. Another advantage to event-related designs is that the effects of fatigue, boredom and systematic patterns of thought unrelated to the task during long inter-trial intervals can be avoided. A drawback is that the power to detect activation is typically lower than for block designs, though the capability to obtain images of more trials per unit time can counter this loss of power.

### Optimized Experimental Designs

What constitutes an optimal experimental design depends on the task, as well as on the ability of the fMRI signal to track changes introduced by the task over time. It also depends on what types of comparisons are of interest. The delay and shape of the BOLD response, scanner drift and physiological noise all conspire to complicate experimental design for fMRI. Not all designs with the same number of trials of a given set of conditions are equal, and the spacing and ordering of events is critical. Some intuitions and tests of design optimality can be gained from a deeper understanding of the statistical analysis of fMRI data.

Several methods have been introduced that allow researchers to optimally select the design parameters, as well as the sequencing of events that should be used in an experiment (Wager and Nichols, 2003; Liu and Frank, 2004). These methods define fitness measures for the estimation efficiency, detection power and randomness of the experiment, and apply search algorithms (e.g., the genetic algorithm) to optimize the design according to the specified criteria. When defining the fitness metrics it is typically assumed that the subsequent data analysis will be performed in the general linear model (GLM) framework described in Section 6.2.1 and that the relationship between stimulus and measured response can be modeled using a linear time invariant system. The use of more complex nonlinear models requires different considerations when defining appropriate metrics, the development of which will be important as such models gain in popularity. Finally, an important consideration relates to assumptions made regarding the shape of the HRF and the noise structure. The inclusion of flexible basis functions and correlated noise into the model will modify the trade-offs between estimation efficiency and detection power, and potentially alter what constitutes an optimal design. Hence, even seemingly minor changes in the model formulation can have a large impact on the efficiency of the design. Together these issues complicate the design of experiments and work remains to find the appropriate balance between them. As research hypotheses ultimately become more complicated, the need for more advanced experimental designs will only increase further and this is clearly an area where statisticians can play an important role.

## 5. PREPROCESSING

Prior to statistical analysis, fMRI data typically undergoes a series of preprocessing steps aimed at removing artifacts and validating model assumptions. The main goals are to minimize the influence of data acquisition and physiological artifacts, to validate statistical assumptions and to standardize the locations of brain regions across subjects in order to achieve increased validity and sensitivity in group analysis. When analyzing fMRI data it is typically assumed that all of the voxels in a particular brain volume were acquired simultaneously. Further, it is assumed that each data point in a specific voxel's time series only consists of a signal from that voxel (i.e., that the participant did not move in between measurements). Finally, when performing group analysis and making population inference, all individual brains are assumed to be registered, so that each voxel is located in the same anatomical region for all subjects. Without preprocessing the data prior to analysis, none of these assumptions would hold and the resulting statistical analysis would be invalid.

The major steps involved in fMRI preprocessing are slice timing correction, realignment, coregistration of structural and functional images, normalization and smoothing. Below each step is discussed in detail.

### Slice Timing Correction

When analyzing 3D fMRI data it is typically assumed that the whole brain is measured simultaneously. In reality, because the brain volume consists of multiple slices that are sampled sequentially, and therefore at different time points, similar time courses from different slices will be temporally shifted relative to one another. Figure 7A illustrates the point. Assume that three voxels contained in three adjacent slices exhibit the same *true* underlying temporal profile. Due to the fact that they are sampled at different time points relative to one another, the corresponding *measured* time courses will appear different. Slice timing correction involves shifting each voxel's time course so that one can assume they were measured simultaneously. This can be done either using interpolation or the Fourier shift theorem to correct for differences in acquisition times.

### Motion Correction

An important issue involved in any fMRI study is the proper handling of any subject movement that may have taken place during data acquisition. Even small amounts of head motion during the course of an experiment can be a major source of error if not treated correctly. When movement occurs, the signal from a specific voxel will be contaminated by the signal from neighboring voxels and the resulting data can be rendered useless. Therefore, it is of great importance to accurately estimate the amount of motion and to use this information to correct the images. If the amount of motion is deemed too severe, it may result in the subject being removed completely from the study.

The first step in correcting for motion is to find the best possible alignment between the input image and some target image (e.g., the first image or the mean image). A rigid body transformation involving 6 variable parameters is used. This allows the input image to be translated (shifted in the $x$, $y$ and $z$ directions) and rotated (altered roll, pitch and yaw) to match the target image. Usually, the matching process is performed by minimizing some cost function (e.g., sums of squared differences) that assesses similarity between the two images. Once the parameters that achieve optimal realignment are determined, the image is resampled using interpolation to create new motion corrected voxel values. This procedure is repeated for each individual brain volume.

### Coregistration and Normalization

Functional MRI data is typically of low spatial resolution and provides relatively little anatomical detail. Therefore, it is common to map the results obtained from functional data onto a high resolution structural MR image for presentation purposes. The process of aligning structural and functional images, called coregistration, is typically performed using either a rigid body (6 parameters) or an affine (12 parameters) transformation.

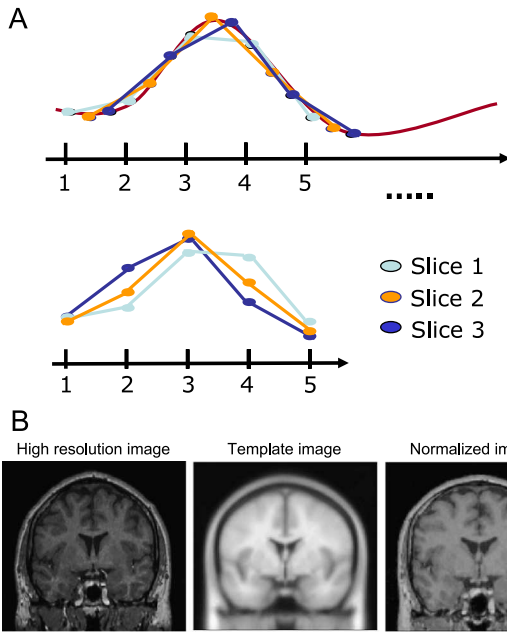

Fig. 7. (A) *Illustration of slice timing correction. Assume three brain slices, exhibiting a similar time course, are sampled sequentially during each TR (top row). Since the voxels are sampled at different time points relative to one another, their respective time courses will appear shifted (bottom row). Slice timing correction shifts the time series so they can be considered to have been measured simultaneously. (B) Illustration of normalization using warping. A high resolution image (left) is warped onto a template image (center), resulting in a normalized image (right).*

For group analysis, it is important that each voxel lie within the same brain structure for each individual subject. Of course individual brains have different shapes and features, but there are regularities shared by every nonpathological brain. Normalization attempts to register each subjects anatomy to a standardized stereotaxic space defined by a template brain [e.g., the Talairach or Montreal Neurological Institute (MNI) brain]. In this scenario using a rigid body transformation is inappropriate due to the inherent differences in the subjects brains. Instead, it is common to use nonlinear transformations to match local features. One begins by estimating a smooth continuous mapping between the points in an input image with those in the target image. Next, the mapping is used to resample the input image so that it is warped onto the target image. Figure 7B illustrates the process, where a high resolution image is warped onto a template image, resulting in a normalized image that can be compared with similarly normalized images obtained from other subjects.

The main benefits of normalizing data are that spatial locations can be reported and interpreted in a consistent manner, results can be generalized to a larger population and results can be compared across studies and subjects. The drawbacks are that it reduces spatial resolution and may introduce errors due to interpolation.

**Spatial Smoothing**

It is common practice to spatially smooth fMRI data prior to analysis. Smoothing typically involves convolving the functional images with a Gaussian kernel, often described by the full width of the kernel at half its maximum height (FWHM). Common values for the kernel widths vary between 4–12 mm FWHM. There are several reasons why it is common to smooth fMRI data. First, it may improve inter-subject registration and overcome limitations in the spatial normalization by blurring any residual anatomical differences. Second, it ensures that the assumptions of random field theory (RFT), commonly used to correct for multiple-comparisons, are valid. A rough estimate of the amount of smoothing required to meet the assumptions of RFT is a FWHM of 3 times the voxel size (e.g., 9 mm for 3 mm voxels). Third, if the spatial extent of a region of interest is larger than the spatial resolution, smoothing may reduce random noise in individual voxels and increase the signal-to-noise ratio within the region.

The process of spatially smoothing an image is equivalent to applying a low-pass filter to the sampled $k$-space data prior to reconstruction. This implies that much of the acquired data is discarded as a byproduct of smoothing and temporal resolution is sacrificed without gaining any benefits. Additionally, acquiring an image with high spatial resolution and thereafter smoothing the image does not lead to the same results as directly acquiring a low resolution image. The signal-to-noise ratio during acquisition increases as the square of the voxel volume, so acquiring small voxels means that signal is lost that can never be recovered. Hence, it is optimal in terms of sensitivity to acquire images at the desired resolution and not employ smoothing. Some recent acquisition schemes have been designed to acquire images at the final functional resolution desired (Lindquist et al., 2008b). This allows for much more rapid image acquisition, as time is not spent acquiring information that will be discarded in the subsequent analysis.

While all the preprocessing steps outlined above are essential for the standard model assumptions required for statistical analysis to hold, there needs to be a clear understanding of the effects they have on both the spatial and temporal correlation structure. More generally, it is necessary to study the interactions among the individual preprocessing steps. For example, is it better to perform slice timing correction or realignment first, and how will this choice impact the resulting data? Ideally there would be one model for both, that also performs outlier detection and correction for physiological noise. There has been increased interest in developing generative models that incorporate multiple steps at once, and this is another problem with a clear statistical component that promises to play an important role in the future.

## 6. DATA ANALYSIS

There are several common objectives in the analysis of fMRI data. These include localizing regions of the brain activated by a certain task, determining distributed networks that correspond to brain function and making predictions about psychological or disease states. All of these objectives are related to understanding how the application of certain stimuli leads to changes in neuronal activity. They are also all intrinsically statistical in nature, and this area is the primary domain of statisticians currently involved in the field. The statistical analysis of fMRI data involves working with massive data sets that exhibit a complicated spatial and temporal noise structure. The size and complexity of the data make it difficult to create a full statistical model for describing its behavior, and a number of shortcuts are required to balance computational feasibility with model efficiency.

### 6.1 Modeling the fMRI Signal

In this section we introduce a generic model for describing fMRI data, and proceed by making a number of model assumptions that impact the direction of the analysis. We begin by assuming that the data consists of a brain volume with $N$ voxels that is repeatedly measured at $T$ different time points. In addition, suppose the experiment is repeated for $M$ subjects. In Section 3.3 the various components present in an fMRI time series were discussed. These included the BOLD response, various nuisance signal and noise. Incorporating all these components, our model for fMRI activation in a single voxel for a single subject can be expressed

$$(3) \quad y_{ij}(t) = \sum_{g=1}^{G} z_{ijg}(t)\gamma_{ijg} + \sum_{k=1}^{K} x_{ijk}(t)\beta_{ijk} + \varepsilon_{ij}(t),$$

for $i = 1, \ldots, N$, $j = 1, \ldots, M$ and $t = 1, \ldots, T$. Here $z_{ijg}(t)$ represents the contribution of nuisance covariates at time $t$, including terms modeling the scanner drift, periodic fluctuations due to heart rate and respiration, and head motion. Similarly, $x_{ijk}(t)$ represents the task-related BOLD response (the signal of interest) corresponding to the $k$th condition at time $t$. The terms $\beta_{ijk}$ and $\gamma_{ijg}$ represent the unknown amplitude of $x_{ijk}$ and $z_{ijg}$, respectively, and $\varepsilon_{ij}(t)$ the noise process. Appropriate ways of modeling each of these signal components are described in detail below.

*The drift component.* In fMRI the signal typically drifts slowly over time due to scanner instabilities. Therefore, most of the power lies in the low-frequency portion of the signal. To remove the effects of drift, it is common to remove fluctuations below a specified frequency cutoff using a high-pass filter. This can be performed either by applying a temporal filter as a preprocessing step, or by including covariates of no interest into the model. As an example of the latter, the drift, $\mu(t)$, can be modeled using a $p$th order polynomial function, that is,

$$(4) \quad \mu_{ij}(t) = \sum_{g=1}^{p} \gamma_{ijg}t^{g-1},$$

which, assuming $z_{ijg}(t) = t^{g-1}$, fits into the framework described in model (3).

There are several alternative functions that have been used to model the drift. For example, it is common to use a series of low frequency cosine functions. The most important issue when using a high-pass filter is to ensure that the fluctuations induced by the task design are not in the range of frequencies removed by the filter, as we do not want to throw out the signal of interest. Hence, the ultimate choice in how to model the drift needs to be made with the experimental design in mind.

*Seasonal components.* Additional covariates may be included to account for periodic noise present in the signal, such as heart-rate and respiration. Physiological noise can in certain circumstances be directly estimated from the data (Lindquist et al., 2008a), or it can be removed using a properly designed band-pass filter. However, in most studies, with TR values ranging from 2–4 s, one cannot hope to estimate and remove the effects of heart-rate and respiration solely by looking at the fMRI time series. Some groups have therefore begun directly measuring heart beat and respiration during scanning and using this information to remove signal related to physiological fluctuations from the data (Glover, Li and Ress, 2000). This is done either as a preprocessing step, or by including these terms as covariates in the model. However, more often than not, the effects of physiological noise are left unmodeled, and the aliased physiological artifacts give rise to the autocorrelated noise present in fMRI data (Lund et al., 2006).

*Noise.* In standard time series analysis, model identification techniques are used to determine the appropriate type and order of a noise process. In fMRI data analysis this approach is not feasible due to the large number of time series being analyzed, and noise models are specified a priori. In our own work, we typically use an auto-regressive process of order 2. The reason we choose an AR model over an ARMA model is that it allows us to use method of moments rather than maximum likelihood procedures to estimate the noise parameters. This significantly speeds-up computation time when repeatedly fitting the model to tens of thousands of time series. Choosing the order of the AR process to be 2 has been empirically determined to provide the most parsimonious model that is able to account for autocorrelation present in the signal due to aliased physiological artifacts.

*The BOLD response.* The relationship between stimuli and BOLD response is typically modeled using a linear time invariant (LTI) system, where the stimulus acts as the input and the HRF acts as the impulse response function. See Figure 4 for an illustration of how the BOLD response varies depending on the stimuli. A linear time invariant system is characterized by the following properties: scaling, superposition and time-invariance. Scaling implies that if the input is scaled by a factor $b$, then the BOLD response will be scaled by the same factor. This is important as it implies that the amplitude of the measured signal provides a measure of the amplitude of neuronal activity. Therefore, the relative difference in amplitude between two conditions can be used to infer that the neuronal activity was similarly different. Superposition implies that the response to two different stimuli applied together is equal to the sum of the individual responses. Finally, time-invariance implies that if a stimulus is shifted by a time $t$, then the response is also shifted by $t$. These three properties allow us to differentiate between responses in various brain regions to multiple closely spaced stimuli.

In our model we allow for $K$ different conditions to be applied throughout the course of the experiment (e.g., varying degrees of painful stimuli). The BOLD response portion of the model can thus be written

$$(5) \qquad s_{ij}(t) = \sum_{k=1}^{K} \beta_{ijk} \int h_{ij}(u)v_k(t-u)\,du,$$

where $h_{ij}(t)$ is the HRF, $v_k(t)$ the stimulus function and $\beta_{ijk}$ the signal amplitude for condition $k$ at voxel $i$ in the $j$th subject.

*Model summary.* For most standard fMRI experiments we can summarize model (3) as

$$y_{ij}(t) = \sum_{g=1}^{p} \gamma_{ijg} t^{g-1}$$

$$(6) \qquad + \sum_{k=1}^{K} \beta_{ijk} \int h_{ij}(u)v_k(t-u)\,du$$

$$+ \varepsilon_{ij}(t),$$

where $\varepsilon_{ij}$ is assumed to follow an AR(2) process. In matrix form this can be written

$$(7) \qquad \mathbf{y}_{ij} = \mathbf{Z}_{ij}\boldsymbol{\gamma}_{ij} + \mathbf{X}_{ij}\boldsymbol{\beta}_{ij} + \boldsymbol{\varepsilon}_{ij},$$

where $\boldsymbol{\gamma}_{ij} = (\gamma_{ij1}, \ldots, \gamma_{ijp})^T$, $\boldsymbol{\beta}_{ij} = (\beta_{ij1}, \ldots, \beta_{ijK})^T$, $\mathbf{Z}_{ij}$ is a $T \times p$ matrix with columns corresponding to the polynomial functions, and $\mathbf{X}_{ij}$ is a $T \times K$ matrix with columns corresponding to the predicted BOLD response for each condition.

Further, the model in (7) can be combined across voxels as follows:

$$(8) \qquad \mathbf{Y}_j = \mathbf{X}_j\mathbf{B}_j + \mathbf{Z}_j\mathbf{G}_j + \mathbf{E}_j.$$

Here $\mathbf{Y}_j$ is a $T \times N$ matrix, where each column is a time series corresponding to a single brain voxel and each row is the collection of voxels that make up an image at a specific time point. The matrices $\mathbf{X}_j$ and $\mathbf{Z}_j$ are the common design matrices used for each voxel. Finally, $\mathbf{B}_j = (\boldsymbol{\beta}_{1j}, \ldots, \boldsymbol{\beta}_{Nj})$, $\mathbf{G}_j = (\boldsymbol{\gamma}_{1j}, \ldots, \boldsymbol{\gamma}_{Nj})$ and $\mathbf{E}_j = (\boldsymbol{\varepsilon}_{1j}, \ldots, \boldsymbol{\varepsilon}_{Nj})$. The vectorized variance of $\mathbf{E}$ is typically assumed to be separable in time and space. In addition, somewhat surprisingly, the spatial covariance is often assumed to be negligible compared to the temporal covariance and therefore ignored.

While (8) provides a framework for a full spatio-temporal model of brain activity, it is currently not considered a feasible alternative due to the extreme computational demands required for model fitting. Instead, model (7) is applied to each voxel separately, and spatial concerns are incorporated at a later stage (see below). Alternatively, the matrix $\mathbf{Y}_j$ is sometimes analyzed using Multivariate methods as described in Section 6.3.

## 6.2 Localizing Brain Activity

The assumptions that one makes regarding the BOLD response fundamentally impact the analysis when using model (7). In most controlled experiments it is reasonable to assume that the stimulus function $v_k(t)$ is known and equivalent to the experimental paradigm (e.g., a vector of zeros and ones where 1 represents time points when the stimulus is "on" and 0 when it is "off"). If one further assumes that the HRF is known a priori, (7) reverts to a multiple regression model with known signal components and unknown amplitudes. These are the assumptions made in the hugely popular GLM approach (Worsley and Friston, 1995; Friston et al., 2002), though the assumption regarding fixed HRF can be relaxed. However, in many areas of psychological inquiry (e.g., emotion and stress), it may be difficult to specify information regarding the stimulus function a priori. If one is unwilling to make any assumptions regarding the exact timing of neuronal activity, alternative methods may be more appropriate for analyzing the data. In the next two sections both scenarios will be discussed.

*6.2.1 The general linear model approach.* The general linear model (GLM) approach has arguably become the dominant way to analyze fMRI data. It models the time series as a linear combination of several different signal components and tests whether activity in a brain region is systematically related to any of these known input functions. The simplest version of the GLM assumes that both the stimulus function and the HRF are known. The stimulus is assumed to be equivalent to the experimental paradigm, while the HRF is modeled using a canonical HRF, typically either a gamma function or the difference between two gamma functions (see Figure 5). Under these assumptions, the convolution term in the BOLD response is a known function and (7) reverts to a standard multiple linear regression model. The BOLD response can be summarized in a design matrix $\mathbf{X}$, containing a separate column for each of the $K$ predictors; see Figure 8 for an example when $K = 2$.

In the remainder of the section we will, for simplicity, assume that the nuisance term $\mathbf{Z}$ is accounted for and can be ignored. Further, we assume a separate, but identical, model for each voxel and suppress the voxel index. Hence, the data for subject $j$ at voxel $i$ can be written

$$(9) \qquad \mathbf{y}_j = \mathbf{X}_j \boldsymbol{\beta}_j + \boldsymbol{\varepsilon}_j,$$

where $\boldsymbol{\varepsilon}_j \sim N(0, \mathbf{V})$ with the structure of the covariance matrix $\mathbf{V}$ corresponding to an AR(2) process with unknown parameters $\phi_1$, $\phi_2$ and $\sigma$. The model parameters can be estimated using a Cochrane–Orcutt fitting procedure, where the variance components are estimated using the Yule–Walker method (Brockwell and Davis, 1998). After fitting the model, one can test for an effect $\mathbf{c}^T \boldsymbol{\beta}_j$ where $\mathbf{c}$ is a contrast vector. The contrast vector can be used to estimate signal magnitudes in response to a single condition, an average over multiple conditions or the difference in magnitude between two conditions. Hypothesis testing is performed in the usual manner by testing individual model parameters using a $t$-test and subsets of parameters using a partial $F$-test. Since the covariance matrix has to be estimated, a Satterthwaite approximation is used to calculate the effective degrees of freedom for the test statistics. This procedure is repeated for brain voxel and the results are summarized in a statistical map consisting of an image whose voxel measurements correspond to the test statistic calculated at that particular voxel.

While the GLM is a simple and powerful approach toward modeling the data, it is also extremely rigid. Even minor mismodeling (e.g., incorrect stimulus function or HRF) can result in severe power loss, and can inflate the false positive rate beyond the nominal value. Due to the massive amount of data, examining the appropriateness of the model is challenging and standard methods of model diagnostics are not feasible. Recently some techniques have been introduced (Luo and Nichols, 2003; Loh, Lindquist and Wager, 2008) that allow one to quickly determine, through graphical representations, areas in the brain where assumptions are violated and model misfit may be present. However, in the vast majority of studies no model checking is performed, calling into question the validity of the results. Moving toward using more sophisticated models, as well as increased use of diagnostics, is an important area of current and future research. In both of these areas statisticians can play an important role.

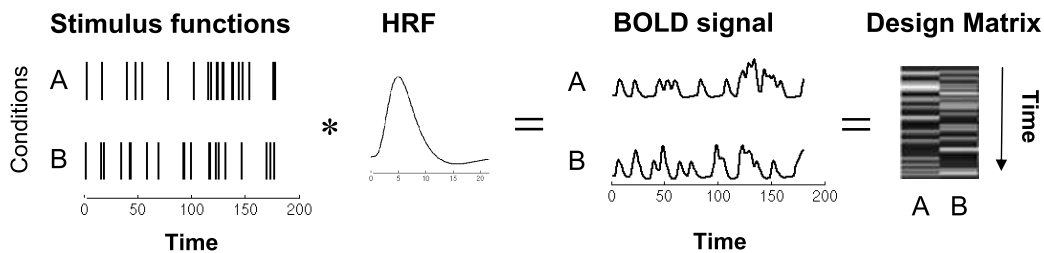

Fig. 8. *In an fMRI experiment with two conditions (A and B), the stimulus function is convolved with a canonical HRF to obtain two sets of predicted BOLD responses. The responses are placed into the columns of a design matrix $\mathbf{X}$ and used to compute whether there is significant signal corresponding to the two conditions in a particular time course.*

As mentioned in Section 3.3, the shape of the HRF may vary across both space and subjects. Therefore, assuming that the shape of the HRF is constant across all voxels and subjects may give rise to significant mismodeling in large parts of the brain. We can relax this assumption by expressing the HRF as a linear combination of reference waveforms. This can be done in the GLM framework by convolving the same stimulus function with multiple canonical waveforms and entering them into multiple columns of $\mathbf{X}$ for each condition. These reference waveforms are called basis functions, and the predictors for an event type constructed using different basis functions can combine linearly to better fit the evoked BOLD responses. The ability of a basis set to capture variations in hemodynamic responses depends both on the number and shape of the reference waveforms. There is a fundamental tradeoff between flexibility to model variations and power, as flexible models can model noise and produce noisier parameter estimates. In addition, the inclusion of additional model parameters decreases the number of degrees of freedom for the subsequent test statistic.

One of the most flexible models, a finite impulse response (FIR) basis set, contains one free parameter for every time-point following stimulation in every cognitive event-type that is modeled (Glover, 1999a; Goutte, Nielsen and Hansen, 2000). Thus, the model is able to estimate an HRF of arbitrary shape for each event type in every voxel of the brain. Another possible choice is to use the canonical HRF together with its temporal derivative in order to allow for small shifts in the onset of the HRF. Other choices of basis sets include those composed of principal components (Aguirre, Zarahn and D'Esposito, 1998; Woolrich, Behrens and Smith, 2004), cosine functions (Zarahn, 2002), radial basis functions (Riera et al., 2004), spectral basis sets (Liao et al., 2002) and inverse logit functions (Lindquist and Wager, 2007b). For a critical evaluation of various basis sets, see Lindquist and Wager (2007b) and Lindquist et al. (2008c).

*Multi-subject analysis.* The analysis so far has been concerned with single subject data. However, researchers typically want to make conclusions on population effects, and statistical analysis needs to be extended to incorporate information from a group of subjects. Multi-subject fMRI data is intrinsically hierarchical in nature, with lower-level observations (e.g., individual subjects) nested within higher levels (e.g., groups of subjects). Multi-level models provide a framework for performing mixed-effects analysis on multi-subject fMRI data. In fMRI it is common to use a two-level model where the first level deals with individual subjects and the second level deals with groups of subjects. In the first-level the data are autocorrelated with a relatively large number of observations, while in the second-level we have IID data with relatively few observations. The first-level model can be written

$$\mathbf{y} = \mathbf{X}\boldsymbol{\beta} + \boldsymbol{\varepsilon}, \tag{10}$$

where $\mathbf{y} = (\mathbf{y}_1^T, \ldots, \mathbf{y}_M^T)^T$, $\mathbf{X} = \mathrm{diag}(\mathbf{X}_1, \ldots, \mathbf{X}_M)$, $\boldsymbol{\beta} = (\boldsymbol{\beta}_1^T, \ldots, \boldsymbol{\beta}_M^T)^T$, $\boldsymbol{\varepsilon} = (\boldsymbol{\varepsilon}_1^T, \ldots, \boldsymbol{\varepsilon}_M^T)^T$ and $\mathrm{Var}(\boldsymbol{\varepsilon}) = \mathbf{V}$ where $\mathbf{V} = \mathrm{diag}(\mathbf{V}_1^T, \ldots, \mathbf{V}_M^T)$.

The second-level model can be written

$$\boldsymbol{\beta} = \mathbf{X}_G\boldsymbol{\beta}_G + \boldsymbol{\varepsilon}_G, \tag{11}$$

where $\boldsymbol{\varepsilon}_G \sim N(0, \mathbf{I}\sigma_G^2)$. Here $\mathbf{X}_G$ is the second-level design matrix (e.g., separating cases from controls) and $\boldsymbol{\beta}_G$ the vector of second-level parameters. The two-level model can be combined into a single level model, which can be expressed as

$$\mathbf{y} = \mathbf{X}\mathbf{X}_G\boldsymbol{\beta}_G + \mathbf{X}\boldsymbol{\varepsilon}_G + \boldsymbol{\varepsilon}. \tag{12}$$

Estimation of the regression parameters and variance components can be performed iteratively, with regression parameters estimated using GLS and variance components estimated using restricted maximum likelihood (ReML) and the EM-algorithm.

Recently, these types of multi-level mixed-effects models have become popular in the neuroimaging community due to their ability to perform valid population level inference (e.g., Friston et al., 2002; Beckmann, Jenkinson and Smith, 2003). However, because of the massive amount of data being analyzed and the fact that it must be feasible to repeatedly fit the model across all brain voxels, the most commonly used techniques are by necessity simplistic. For example, they do not readily allow for unbalanced designs and missing data. However, both issues are prevalent in fMRI data analysis. Missing data may be present in a study because of artifacts and errors due to the complexity of data acquisition (including human error), while unbalanced designs are important because of interest in relating brain activity to performance and other variables that cannot be experimentally controlled. The introduction of techniques for performing rapid estimation of multi-level model parameters that allow for this type of data is of utmost importance. Multilevel models have been heavily researched in the statistical community, and statisticians can play an important role in developing methods tailored directly to the complexities of fMRI data analysis.

*Spatial modeling.* Up to this point the entire analysis procedure outlined in this section has been univariate, that is, performed separately at each voxel. Indeed, one of the most common short cuts used in the field is, somewhat surprisingly, to perform fMRI data analysis in a univariate setting (the so-called "massive univariate approach"), where each voxel is modeled and processed independently of the others. At the model-level this approach assumes that neighboring voxels are independent, which is generally not a reasonable assumption as most activation maps show a clear spatial coherence. In these situations the spatial relationship is sometimes accounted for indirectly by smoothing the data prior to voxel-wise analysis, and thereafter applying random field theory to the map of test statistics to determine statistical significance for the entire set of voxels. Hence, the "massive univariate approach" does take spatial correlation into account at the level of thresholding using Gaussian random fields. However, while the random field theory approach does link voxel-wise statistics, it does not directly estimate spatial covariances under a linear model. We discuss random field theory further in Section 6.2.3.

Incorporating spatial considerations into the GLM framework has become a subject of increased interest in recent years. In the earliest approaches individual voxel-wise GLMs were augmented with time series from neighboring voxels (Katanoda, Matsuda and Sugishita, 2002; Gossl, Auer and Fahrmeir, 2001). Recently, a series of Bayesian approaches have been suggested. Penny, Trujillo-Barreto and Friston (2005) have proposed a fully Bayesian model with spatial priors defined over the coefficients of the GLM. Bowman (Bowman, 2005) presents a whole-brain spatio-temporal model that partitions voxels into functionally related networks and applies a spatial simultaneous autoregressive model to capture intraregional correlations between voxels. Finally, Woolrich et al. (2005) have developed a spatial mixture model using a discrete Markov random field (MRF) prior on a spatial map of classification labels. While these models are certainly a step in the right direction, it is clear that the massive univariate approach continues to be exceedingly popular among end users due to its relative simplicity.

Some headway has recently been made, but work remains to be done and ideas from spatial statistics can potentially play an important role. Fitting spatial models using Bayesian statistics has been the focus of much attention lately and several promising approaches have been suggested (e.g., Bowman, 2005; Bowman et al., 2008; Woolrich et al., 2005). However, model complexity is sometimes constrained by the massive amounts of data and there is a clear need for statisticians with strong training in Bayesian computation to optimize the model fitting procedure.

6.2.2 *Data with uncertain timing of activation.* In many areas of psychological inquiry—including studies on memory, motivation and emotion—it is hard to specify the exact timing of activation a priori. In this situation it may not be reasonable to assume that either the experimental paradigm or the HRF are known. Therefore, the GLM cannot be directly applied to these data sets and alternative methods are needed. Typically, researchers take a more data-driven approach that attempts to characterize reliable patterns in the data, and relate those patterns to psychological activity post hoc. One popular approach is independent components analysis (ICA) (Beckmann and Smith, 2005; Calhoun et al., 2001b; McKeown and Makeig, 1998), a member of a family of analytic methods that also includes principal components and factor analysis. While these methods provide a great deal of flexibility, they do not provide a formal framework for performing inference about whether a component varies over time and when changes occur in the time series. In addition, because they do not contain any model information, they capture regularities whatever the source. Therefore, they are highly susceptible to noise and components are often dominated by artifacts. For these reasons we prefer to use methods from change point analysis to model fMRI data with unknown activation profiles.

In our own work, we use a three step procedure for modeling such data. In a first stage we employ a multi-subject (mixed-effects) extension of the exponentially weighted moving average (EWMA) method (Roberts, 1959), denoted HEWMA (Hierarchical EWMA) (Lindquist and Wager, 2007a), as a simple screening procedure to determine which voxels have time courses that deviate from a baseline level and should be moved into the next stage of the analysis. In the second stage we estimate voxel-specific distributions of onset times and durations from the fMRI response, by modeling each subject's onset and duration as random variables drawn from an unknown population distribution (Robinson, Wager and Lindquist, 2009). We estimate these distributions assuming no functional form (e.g., no assumed neural or hemodynamic response), and allowing for the possibility that some subjects may show no response. The distributions can be used to estimate the probability that a voxel is activated as a function of time. In the final step we perform spatial clustering of voxels according to onset and duration characteristics, and anatomical location using a hidden Markov random field model (Robinson, Wager and Lindquist, 2009). This three step procedure provides a spatio-temporal model for dealing with data with uncertain onset and duration.

There exists a rich literature on sequential and change point analysis with applications to a wide range of fields. However, to date there have been relatively few applications of these methods to fMRI data. As experimental paradigms and the psychological questions researchers seek to understand become more complicated, these methods could possibly play an important role. Therefore, this is an area where statisticians can make a contribution.

### 6.2.3 *Multiple comparisons.*

The results of fMRI studies are usually summarized in a statistical parametric map (SPM), such as the one shown in Figure 9. These maps describe brain activation by color-coding voxels whose *t*-values (or comparable statistics) exceed a certain statistical threshold for significance. The implication is that these voxels are activated by the experimental task. When constructing such a map it is important to carefully consider the appropriate threshold to use when declaring a voxel active. In a typical experiment up to 100,000 hypothesis tests (one for each voxel) are performed simultaneously, and it is crucial to correct for multiple comparisons. Several approaches toward controlling the false positive rate have been used; the fundamental difference between methods lies in whether they control the family-wise error rate (FWER) or the false discovery rate (FDR).

Random Field Theory (RFT) (Worsley et al., 2004) is the most popular approach for controlling the FWER in the fMRI community. Here, the image of voxel-wise test statistic values are assumed to be a discrete sampling of a continuous smooth random field. In the RFT approach one begins by estimating the smoothness of the image, which is expressed in terms of resolution elements, or resels (roughly equivalent to the number of independent comparisons). Next, using information about the number of resels and the shape of the search volume, mathematical theory exists for calculating the expected Euler characteristic. For large thresholds this value is equal to the number of clusters of activity that one would expect by chance at a certain statistical threshold. Hence, it can be used to determine the appropriate threshold that controls the FWER at a certain level. RFT is a mathematically elegant approach toward correcting for multiple comparisons. However, like most other methods that control the FWER, it tends to give overly conservative results (Hayasaka and Nichols, 2004). If one is unwilling to make assumptions about the distribution of the data, nonparametric methods can be used to control the FWER. It has been shown that such methods can provide substantial improvements in power and validity, particularly with small sample sizes (Nichols and Holmes, 2002).

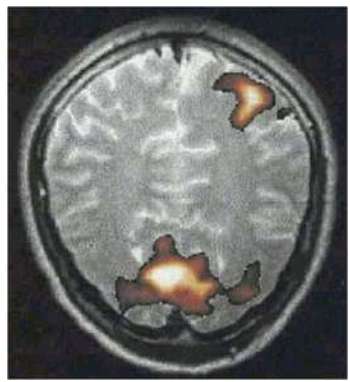

FIG. 9. *Statistical parametric maps (SPM) are used to present the results of the statistical analysis. Voxels whose p-values are below a certain threshold are color-coded to signify that they contain significant task-related signal. The results are superimposed onto a high-resolution anatomical image for presentation purposes.*

The false discovery rate (FDR) controls the proportion of false positives among all rejected tests and has recently been introduced to the neuroimaging community (Genovese, Lazar and Nichols, 2002). The FDR controlling procedure is adaptive in the sense that the larger the signal, the lower the threshold. If all of the null hypotheses are true, the FDR will be equivalent to the FWER. Any procedure that controls the FWER will also control the FDR. Hence, any procedure that controls the FDR can only be less stringent and lead to increased power. A major advantage is that since FDR controlling procedures work only on the $p$-values and not on the actual test statistics, it can be applied to any valid statistical test. In contrast, for the RFT approach the test statistics need to follow a known distribution.

The FDR controlling procedure that is most commonly used in fMRI data analysis is the so-called Benjamini–Hochberg procedure (Benjamini and Hochberg, 1995), where all tests are assumed to be independent. However, in fMRI data analysis it is more realistic to assume that tests are dependent, as neighboring voxels are more likely to have similarly valued $p$-values. Hence, the introduction of FDR controlling procedures that incorporate spatial information is of utmost importance and an area of future research for statisticians.

### 6.3 Assessing Brain Connectivity

Human brain mapping has primarily been used to construct maps indicating regions of the brain that are activated by specific tasks. Recently, there has been an increased interest in augmenting this type of analysis with connectivity studies that describe how various brain regions interact and how these interactions depend on experimental conditions. It is common in the fMRI literature to distinguish between anatomical, functional and effective connectivity (Friston, 1994). Anatomical connectivity deals with describing how different brain regions are physically connected and can be tackled using diffusion tensor imaging (DTI). Functional connectivity is defined as the undirected association between two or more fMRI time series, while effective connectivity is the directed influence of one brain region on others. In this work we concentrate on describing the latter two types of connectivity.

6.3.1 *Functional connectivity.* The simplest approach toward functional connectivity analyses compares correlations between regions of interest, or between a "seed" region and other voxels throughout the brain. Alternative approaches include using multivariate methods, such as Principal Components Analysis (PCA) (Andersen, Gash and Avison, 1999) and Independent Components Analysis (ICA) (Calhoun et al., 2001b; McKeown and Makeig, 1998), to identify task-related patterns of brain activation without making any a priori assumptions about its form. These methods involve decomposing the $T \times$ $N$ data matrix, $\mathbf{Y}$, into a set of spatial and temporal components according to some criteria.[2]

PCA allows one to determine the spatial patterns that account for the greatest amount of variability in a time series of images. This can be achieved by finding the singular value decomposition of the data matrix,

$$(13) \qquad \mathbf{Y} = \mathbf{U}\mathbf{S}\mathbf{V}^{T},$$

where $\mathbf{U}$ is an $T \times T$ unitary matrix, $\mathbf{S}$ is a $T \times N$ diagonal matrix with nonnegative elements, and $\mathbf{V}$ is an $N \times N$ unitary matrix. The columns of $\mathbf{U}$ represent the weighted sum of time series from different voxels, while the columns of $\mathbf{V}$ contain the voxel weights required to create each component in $\mathbf{U}$. Thus, $\mathbf{U}$ represents the temporal components and $\mathbf{V}$ the spatial components of the data. The elements of $\mathbf{S}$ represent the amount of variability explained by each component and are ordered in nonincreasing fashion. Hence, the first column of $\mathbf{V}$ shows how to weight each of the $N$ voxel time series in order to capture the most variance in $\mathbf{Y}$, etc. The usefulness of this technique is twofold: this decomposition can potentially reveal the nature of the observed signal by finding its linearly independent sources. Also, decomposing the signal and ordering the components according to their weight is a useful way to simplify the data or filter out unwanted components, and can be used in the preprocessing stage as a data reduction tool.

ICA is similar to PCA, but the components are required to be independent rather than orthogonal. ICA assumes that $\mathbf{Y}$ is a weighted sum of $p$ independent source signals contained in the $p \times N$ source matrix $\mathbf{X}$, whose weights are described by a $T \times p$ mixing matrix of weights $\mathbf{M}$, that is,

$$(14) \qquad \mathbf{Y} = \mathbf{M}\mathbf{X}.$$

Iterative search algorithms are used to estimate $\mathbf{M}$ and $\mathbf{X}$, simultaneously. In order to solve (14), ICA makes a number of assumptions, the main ones being that the data consist of $p$ statistically independent components, where at most one component is Gaussian. The independence assumption entails that the activations do not have a systematic overlap in time or space, while the non-Gaussiantity assumption is required for the problem to be well defined. An ICA of $\mathbf{Y}$ produces spatially independent component images in the matrix $\mathbf{X}$ and is usually referred to as spatial ICA (sICA). Performing ICA on $\mathbf{Y}^T$ instead produces temporally independent component time series and is referred to as temporal ICA (tICA).

---

[2]Note that throughout this section we have suppressed the subject index previously used.

Both PCA and ICA reduce the data to a lower-dimensional space by capturing the most prominent variations across the set of voxels. The components may either reflect signals of interest or they may be dominated by artifacts; it is up to the user to determine which are important. Both ICA and PCA assume all variability results from the signal, as noise is not included in the model formulation, though noise-added versions of ICA that account for non-source noise have been introduced (Hyvarinen, Karhunen and Oja, 2001). In ICA, interpretation is made more difficult by the fact that the sign of the independent components cannot be determined. In addition, the independent components are not ranked in order of appearance and it is therefore necessary to sift through all of the components to search for the ones that are important.

ICA has been successfully applied to single-subject fMRI data. Extending the approach to allow for group inference is currently an active area of research. Several techniques for performing multisubject ICA have been proposed. The GIFT approach (Calhoun et al., 2001a) consists of temporally concatenating the data across subjects, and performing ICA decomposition on the resulting data set. In contrast, the tensor ICA (Beckmann and Smith, 2005) approach factors multisubject data as a trilinear combination of three outer products. This results in a three-way decomposition that represents the data in terms of their temporal, spatial and subject-dependent variations. Finally, Guo and Pagnoni (Guo and Pagnoni, 2008) have proposed a unified framework for fitting group ICA models. They consider a class of models, assuming independence in the spatial domain, which incorporate existing methods such as the GIFT and tensor ICA as special cases. In general, the ability to perform functional connectivity analysis in the multisubject domain promises to be an area of intense research in the future.

6.3.2 *Effective connectivity.* In effective connectivity analysis a small set of regions with a proposed set of directed connections are specified a priori, and tests are used to assess the statistical significance of individual connections. Most effective connectivity methods depend on two models: a neuroanatomical model that describes the areas of interest, and a model that describes how they are connected. Commonly used methods include Structural Equation Modeling (SEM) (McIntosh and Gonzalez-Lima, 1994) and Dynamic Causal Modeling (DCM) (Friston, Harrison and Penny, 2003).

In SEM the emphasis lies on explaining the variance-covariance structure of the data. It comprises a set of regions and directed connections between them. Further, path coefficients are defined for each link representing the expected change in activity of one region given a unit change in the region influencing it. The path coefficient indicates the average influence across the time interval measured. Algebraically, we can express an SEM model as

$$Y = MY + \varepsilon, \tag{15}$$

where $\mathbf{Y}$ is the data matrix, $\mathbf{M}$ is a matrix of path coefficients and $\boldsymbol{\varepsilon}$ is independent and identically distributed Gaussian noise. This can be rewritten

$$Y = (I - M)^{-1}\varepsilon, \tag{16}$$

where $\mathbf{I}$ represents the identity matrix. The solution of the unknown coefficients contained in $\mathbf{M}$ is obtained by studying the empirical covariance matrix of $\mathbf{Y}$. In SEM we seek to minimize the difference between the observed covariance matrix and the one implied by the structure of the model (16). The parameters of the model are adjusted iteratively to minimize the difference between the observed and modeled covariance matrix. All inference rests on the use of nested models and the likelihood ratio test (LRT) to determine whether a path coefficient is reliably different from zero.

A number of model assumptions are made when formulating the SEM. The data are assumed to be normally distributed and independent from sample to sample. An important consequence of this assumption is that SEM discounts temporal information. Consequently, permuted data sets produce the same path coefficients as the original data, which is a major weakness, as the assumption of independence is clearly violated in the analysis of a single subject.

The measurements used in SEM analysis are based on the observed BOLD response and this ultimately limits the scope of any interpretation that can be made at the neuronal level. Dynamic Casual Modeling (DCM) (Friston, Harrison and Penny, 2003) is an attempt to move the analysis to the neuronal level.

DCM uses a standard state-space design, and treats the brain as a deterministic nonlinear dynamic system that is subject to inputs and produces outputs. It is based on a neuronal model of interacting cortical regions, supplemented with a forward model describing how neuronal activity is transformed into measured hemodynamic response. Effective connectivity is parameterized in terms of the coupling among unobserved neuronal activity in different regions. We can estimate these parameters by perturbing the system and measuring the response. Experimental inputs cause changes in effective connectivity at the neuronal level which, in turn, causes changes in the observed hemodynamics.

DCM uses a bilinear model for the neuronal level and an extended Balloon model (Buxton, Wong and Frank, 1998) for the hemodynamic level. In a DCM model the user specifies a set of experimental inputs (the stimuli) and a set of outputs (the activity in each region). The task of the algorithm is then to estimate the parameters of the system, or the "state variables." Each region has five state variables, four corresponding to the hemodynamic model and a fifth corresponding to neuronal activity. The estimation process is then carried out using Bayesian methods, where Normal priors are placed on the model parameters and an optimization scheme is used to estimate parameters that maximize the posterior probability. The posterior density is then used to make inferences about the significance of the connections between various brain regions.

While many researchers use SEM and DCM in order to ascribe causality between different brain regions, it is important to keep in mind that the tests performed by both techniques are based on model fit rather than on the causality of the effect. Any misspecification of the underlying model can lead to erroneous conclusions. In particular, the exclusion of important lurking variables (e.g., brain regions involved in the network but not included in the model) can completely change the fit of the model and thereby affect both the direction and strength of the connections. Therefore, a great deal of care needs to be taken when interpreting the results of these methods.

Granger causality (Roebroeck, Formisano and Goebel, 2005) is another approach that is considered to test effective connectivity. This approach does not rely on a priori specification of a structural model, but rather quantifies the usefulness of past values from one brain region in predicting current values in another. Granger causality provides information about the temporal precedence of relationships among two regions, but is a misnomer because it does not actually provide information about causality.

Let $\mathbf{x}$ and $\mathbf{y}$ be two time courses of length $T$ extracted from two voxels. First, each time course is modeled using a linear autoregressive model of the $p$th order (where $p \leq T - 1$). Second, each time course model is expanded using the autoregressive terms from the other voxel, that is,

$$
\begin{aligned}
x(n) = {} & \sum_{i=1}^{p} a(i)x(n-i) \\
& + \sum_{i=1}^{p} b(i)y(n-i) + \varepsilon_x(n),
\end{aligned}
\tag{17}
$$

$$
\begin{aligned}
y(n) = {} & \sum_{i=1}^{p} a(i)y(n-i) \\
& + \sum_{i=1}^{p} b(i)x(n-i) + \varepsilon_y(n),
\end{aligned}
\tag{18}
$$

where both $\varepsilon_x$ and $\varepsilon_y$ are defined to be white noise processes. In this formulation the current value of both time courses are assumed to depend both on the past $p$-values of its own time course as well as the past $p$-values of the other time course. By fitting both of these models, one can test whether the previous history of $\mathbf{x}$ has predictive value on the time course $\mathbf{y}$ (and vice versa). If the model fit is significantly improved by the inclusion of the cross-autoregressive terms, it provides evidence that the history of one of the time courses can be used to predict the current value of the other and a Granger-causal relationship is inferred.

While the analysis of brain connectivity has been an area of intense research the past couple of years, it has primarily been concerned with analyzing connectivity between different brain regions. However, there is increasing interest in studying networks that incorporate information about performance scores on the task and/or physiological measures. For example, it may be of interest to determine brain regions that mediate changes in heart rate or increases in reported stress in response to a task (Wager et al., 2008). The incorporation of this information is complicated by the fact that the different components included in the network measure different types of responses, possibly on completely different time scales. These types of extensions of current connectivity methods are an area where statisticians can play an important role in the future.

6.3.3 *Understanding connectivity.* Ultimately, the distinction between functional and effective connectivity is not entirely clear (Horwitz, 2003). If the discriminating features are a directional model in which causal influences are specified and the ability to draw conclusions about direct vs. indirect connections, then many analyses (e.g., multiple regression) might count as effective connectivity. In the end, it is not the label that is important, but the specific assumptions and robustness and validity of inference afforded by each method. When performing connectivity studies researchers are often interested in making statements regarding causal links between different brain regions. However, the idea of causality is a very deep and important philosophical issue (Rubin, 1974; Pearl, 2000). Often a cavalier attitude is taken in attributing causal effects and the differentiation between explanation and causation is often blurred. Properly randomized experimental designs permit causal inferences of task manipulations on brain activity. However, in fMRI studies, all the brain variables are observed, and none are manipulated. It is therefore difficult to make strong conclusions about causality and direct influences among brain regions, because the validity of such conclusions is difficult to verify. In general, the area of brain connectivity is experiencing certain growing pains. There is a clear need for a discussion of the goals of the analysis, as well as which model assumptions are reasonable. To date, many of these critical issues have not been properly addressed, and terms such as causality are used inappropriately. In addition, there is also room for introducing new techniques for testing connectivity and ultimately we believe ideas from casual inference will come to play a role.

## 7. ADDITIONAL OPEN STATISTICAL CHALLENGES

Throughout this paper we have attempted to highlight the many interesting and important statistical problems that arise in fMRI research. It is clear that analyzing these massive data structures with their complex correlation patterns provides a serious challenge for researchers in the field. Many standard statistical techniques are neither appropriate nor feasible for direct application to fMRI data. As experimental designs and imaging techniques become more sophisticated, the need for novel statistical methodology will only increase. As we look toward the future, there are many open statistical challenges that need to be addressed for fMRI to reach its full potential. We have attempted to highlight many of these challenges throughout, but below we discuss several additional topics in detail.

### Classification and Prediction

There is a growing interest in using fMRI data as a tool for classification of mental disorders, brain-based nosology and predicting the early onset of disease. For example, the promise of using fMRI as a screening device in detecting early onset of Alzheimer's disease holds obvious appeal. In addition, there has been growing interest in developing methods for predicting stimuli directly from functional data. This would allow for the possibility to infer information from the scans about the subjects thought process and use brain activation patterns to characterize subjective human experience. A particularly controversial application has been the idea of using fMRI for lie detection. The efficient prediction of brain states is a challenging process that requires the application of novel statistical and machine learning techniques. Various multivariate pattern classification approaches have successfully been applied to fMRI data in which a classifier is trained to discriminate between different brain states and then used to predict the brain states in a new set of fMRI data. To date, efficient preprocessing of the data has been shown to be more important than the actual method of prediction. However, this is an area that without a doubt will be the focus of intense research in the future and where statisticians are well positioned to make a significant impact.

### Multi-modal Techniques

In neuroscience there is a general trend toward using multiple imaging methods in tandem to overcome some of the limitations of each method used in isolation. For example, recent advances in engineering and signal processing allow electroencephalography (EEG) and fMRI data to be collected simultaneously (Goldman et al., 2000). EEG has an extremely high temporal resolution (on the order of $ms$) but poor spatial resolution, while fMRI suffers from the opposite problem. By merging these two techniques, the hope is that one can get the best of both worlds. In another example, diffusion tensor imaging (DTI), a popular technique for measuring directional diffusion and reconstructing the fiber tracts of the brain (Le Bihan et al., 2001), can be combined with fMRI to determine appropriate regions of the brain to include in subsequent connectivity models. Finally, neuroimaging data are being combined with transcranial magnetic stimulation (TMS) to integrate the ability of neuroimaging to observe brain activity with the ability of TMS to manipulate brain function (Bohning et al., 1997). Using this technique, one can simulate temporary "brain lesions" while the subject performs certain tasks. One can then attempt to infer causal relationships by studying differences in a brain network when a region is functioning and when it is not.

Combining information from different modalities will be challenging to data analysts, if for no other reason than that the amount of data will significantly increase. In addition, since the different modalities are measuring fundamentally different quantities, it is not immediately clear how to best combine the information. This is an extremely important problem that has already started to become a major area of research.

### Imaging Genetics

The past several decades have seen rapid advances in the study of human brain function. But perhaps even more impressive have been the advances in molecular genetics research that have taken place in the same time period. However, despite the enormous amount of research performed in both of these areas, relatively little work has been done on combining these two types of data.

Integrating genetics with brain imaging is an important problem that has the potential to fundamentally change our understanding of human brain function in diseased populations. It could provide a way to study how a particular subset of polymorphisms affects functional brain activity. In addition, quantitative indicators of brain function could facilitate the identification of the genetic determinants of complex brain-related disorders such as autism, dementia and schizophrenia (Glahn, Thompson and Blangero, 2007). These indicators may also aid in gene discovery and help researchers understand the functional consequences of specific genes at the level of systems neuroscience. Imaging genetics promises to be an important topic of future research, and to fully realize its promise, novel statistical techniques will be needed.

The open statistical challenges discussed in this paper are by no means complete. Rather, we hope that they illustrate some of the possible statistical problems that may be at the forefront of the statistical analysis of fMRI data in the future. Other problems that will be of importance include the acquisition and analysis of real-time fMRI data, the development of efficient nonlinear models for describing the relationship between stimulus and BOLD response, and the synthesis of findings across studies (e.g., meta-analyses), among many other things.

A critical job for any statistician involved in the field will be stressing the need for researchers to stringently state and check model assumptions. Due to the enormity of the analysis, model assumptions are typically neither checked nor often even specified. However, for most models even relatively small amounts of mismodeling can result in severe power loss, and inflate the false positive rate beyond the nominal value. As inference may be incorrect if model assumptions do not hold, the lack of diagnostics calls some of the validity of the analysis into doubt. This is an area where statisticians must lead the way.

## 8. CONCLUSIONS

There has been explosive interest in the use of fMRI in recent years. The rapid pace of development and the interdisciplinary nature of the neurosciences present an enormous challenge to researchers. Moving the field forward requires a collaborative team with expertise in psychology, neuroanatomy, neurophysiology, physics, biomedical engineering, statistics, signal processing and a variety of other disciplines depending on the research question. True interdisciplinary collaboration is exceedingly challenging, because team members must know enough about the other disciplines to be able to talk intelligently with experts in each field. Due to the importance that statistics plays in this research, it is important that more statisticians get involved in these research teams for the methodology to reach its full potential. Through the course of this paper, we have attempted to illustrate that many of the problems involved in studying these complicated data structures are intrinsically statistical in nature. As the experimental design and imaging techniques become more sophisticated, the need for novel statistical methodology will only increase, promising an exciting future for statisticians in the field.